\documentclass[apj]{emulateapj}
\usepackage{graphicx}
\usepackage{epsfig,psfrag}
\usepackage{pslatex}

\shorttitle{Quasar Feedback from Cross-Correlation}
\shortauthors{Chatterjee et al.}

\begin{document}
\title{Tentative Detection of Quasar Feedback from WMAP and SDSS Cross-Correlation} 
\author{Suchetana Chatterjee$^{1,4}$, Shirley Ho$^{2,3}$, Jeffrey A. Newman$^{1}$, and Arthur Kosowsky$^{1}$ \\
$^{1}$Department of Physics and Astronomy, University of Pittsburgh, Pittsburgh, PA 15260 USA\\
$^{2}$Department of Astrophysical Sciences, Princeton University, Princeton, NJ 08544-1001 USA\\
$^{3}$Lawrence Berkeley National Lab, Berkeley, CA, 94720 USA\\
$^{4}$Department of Astronomy, Yale University, New Haven, CT-06520-8101 USA\\}

\begin{abstract}
We perform a cross-correlation analysis of microwave data from Wilkinson Microwave Anisotropy Probe and photometric quasars from the Sloan Digital Sky Survey, testing for Sunyaev-Zeldovich (SZ) effect from quasars. 
A statistically significant (2.5 $\sigma$) temperature decrement exists in the 41 GHz microwave band. A two-component fit to the cross-correlation spectrum incorporating both dust emission and SZ yields a best-fit $y$ parameter of $(7.0 \pm 3.4)\times 10^{-7}$. A similar cross-correlation analysis with the Luminous Red Galaxy sample from Sloan gives a best-fit $y$ parameter of $(5.3 \pm 2.5)\times 10^{-7}$. We discuss the possible physical origin of these signals, which is likely a combination of SZ effects from quasars and from galaxy clusters. Both the Planck Surveyor satellite and current ground-based arcminute-resolution microwave experiments will detect this
signal with a higher statistical significance.    
\end{abstract}

\section{Introduction}
Secondary anisotropies in the cosmic microwave background (CMB) provide a number of opportunities for probing the growth of structure in the Universe. Numerous recent and current experiments aim to detect these effects. 
The thermal Sunyaev-Zeldovich effect (TSZ) (Sunyaev \& Zeldovich 1972), which is the inverse Compton scattering of cosmic microwave background (CMB) photons due to hot electrons, creates a spectral distortion in the CMB and is the dominant secondary effect contributing to arcminute-scale anisotropies. The TSZ effect arises from the random thermal motion of the electrons. If there is a finite velocity of the cluster in the CMB frame there will be an additional Doppler term. This Doppler anisotropy is called the kinetic SZ (KSZ) distortion. The TSZ effect serves as a powerful tool for probing accumulations of hot gas (see Carlstrom, Holder, \& Reese 2002 for a review). Most of  the TSZ signal comes from galaxy clusters, which contain the majority of the thermal energy of the Universe. State-of-the-art SZ surveys like the Atacama Cosmology Telescope (ACT\footnote[1]{http://www.physics.princeton.edu/act/}) (Hincks et al.\ 2009) and the South Pole Telescope (SPT\footnote[2]{http://pole.uchicago.edu}) (Staniszewski et al.\ 2009) aim to detect large numbers of clusters in blind surveys via their SZ signatures. 

A number of other astrophysical processes will also create both TSZ and KSZ distortions. These include KSZ distortions from peculiar velocities during reionization (McQuinn et al. 2005, Illiev et al. 2006), both thermal and kinetic distortions from supernova-driven galactic winds (White, Hernquist, \& Springel 2002; Majumdar, Nath, \& Chiba 2001), thermal effect from black hole-seeded proto-galaxies (Aghanim, Balland, \& Silk 2000; de Zotti et al. 2004, Rosa-Gonz'alez et al.\ 2004, Massardi et al.\ 2008), KSZ from Lyman-break galaxy outflows (Babich \& Loeb 2007), and thermal SZ effect from supernovae from the first generation of stars (Oh, Cooray, \& Kamionkowski 2003). Here we consider another generic signal: the TSZ distortion from hot gas surrounding active galactic nuclei (AGN) (e.g., Natarajan \& Sigurdsson 1999; Yamada, Sugiyama, \& Silk 1999; Lapi, Cavaliere, \& De Zotti 2003; Platania et al.\ 2002; Roychowdhury, Ruszkowski, \& Nath 2005; Chatterjee \& Kosowsky 2007; Scannapieco, Thacker, \& Couchman 2008; Chatterjee et al.\ 2008). The SZ effect from these sources provides a new observational tool to study the role of baryonic physics in structure formation. 
    
X-ray evidence of the effect of AGN feedback on structure formation at different scales has been shown by several groups (e.g., Mcnamara et al.\ 2005; Voit \& Donahue 2005; Sanderson, Ponman, \& O'Sullivan 2006; Weinmann et al.\ 2006; Schawinski et al.\ 2007). At the same time, study of radio jets has illuminated the morphology and outburst history of AGNs (e.g., Birzan et al.\ 2004; Dunn, Fabian, \& Taylor 2005; Dunn \& Fabian 2006). Although effects of AGN feedback have been successfully detected on different scales, the challenge lies in quantifying the amount of feedback energy and its particular effects on structure formation. Adding SZ (thermal from now on) measurements to X-ray data will provide stronger constraints on feedback effects than X-ray data alone (e.g., Scannapieco, Thacker, \& Couchman 2008; Chatterjee et al.\ 2008; Moodley et al.\ 2008). But the AGN signal is tiny, and direct detection through pointed observations with, for example, the Atacama Large Millimeter Array (ALMA\footnote[3]{http://www.alma.nrao.edu/}) (Lapi, Cavaliere, \& De Zotti 2003; Chatterjee \& Kosowsky 2007) will be challenging. In this work we instead cross-correlate CMB maps with optical quasar catalogs (Chatterjee \& Kosowsky 2007, Scannapieco, Thacker, \& Couchman 2008) to enhance the detectability
of the SZ signal.
 
The cross-correlation function and its Fourier transform, the cross-power spectrum (Peebles 1980), have been powerful techniques in cosmology to probe small signals. The cross-correlation of CMB data sets with galaxy surveys has proven useful for measuring secondary anisotropies in the CMB (e.g., Peiris \& Spergel 2000; Refregier, Spergel, \& Herbig 2000; Fosalba, Gaztanaga, \& Castander 2003; Afshordi, Loh, \& Strauss 2003; Padmanabhan et al.\ 2005; Ho et al.\ 2008; Giannantonio et al.\ 2008; Hirata et al.\ 2004; Smith, Zahn, \& Dore 2007; Diego, Silk, \& Sliwa 2003; Cheng, Wu, \& Cooray 2006; Croft, Banday, \& Hernquist 2006; Myers et al.\ 2004; Kashlinsky et al.\ 2009; Diego \& Partridge 2010; Hernandez-Monteagudo et al.\ 2006; Ho, Dedeo, \& Spergel 2009). Here we investigate the SZ distortion from AGN feedback by cross-correlating the WMAP CMB maps with a photometric quasar catalog selected from the Sloan Digital Sky Survey (SDSS).
 
Our paper is organized as follows. In \S 2 we give a brief description of the data sets used in this work. Section 3 discusses our methodology for estimating the amplitude of the SZ-quasar cross-correlation. In \S 4 we describe potential systematic effects and the masks used to eliminate data which is likely to be dominated by systematic errors. Our results are discussed in \S 5. Finally, in \S 6 we give a brief summary of our results and discuss the prospects of detecting quasar-SZ cross-correlations with future experiments.

\begin{figure*}[t]
\begin{center}
\resizebox{80mm}{!}{\includegraphics[angle=90.0]{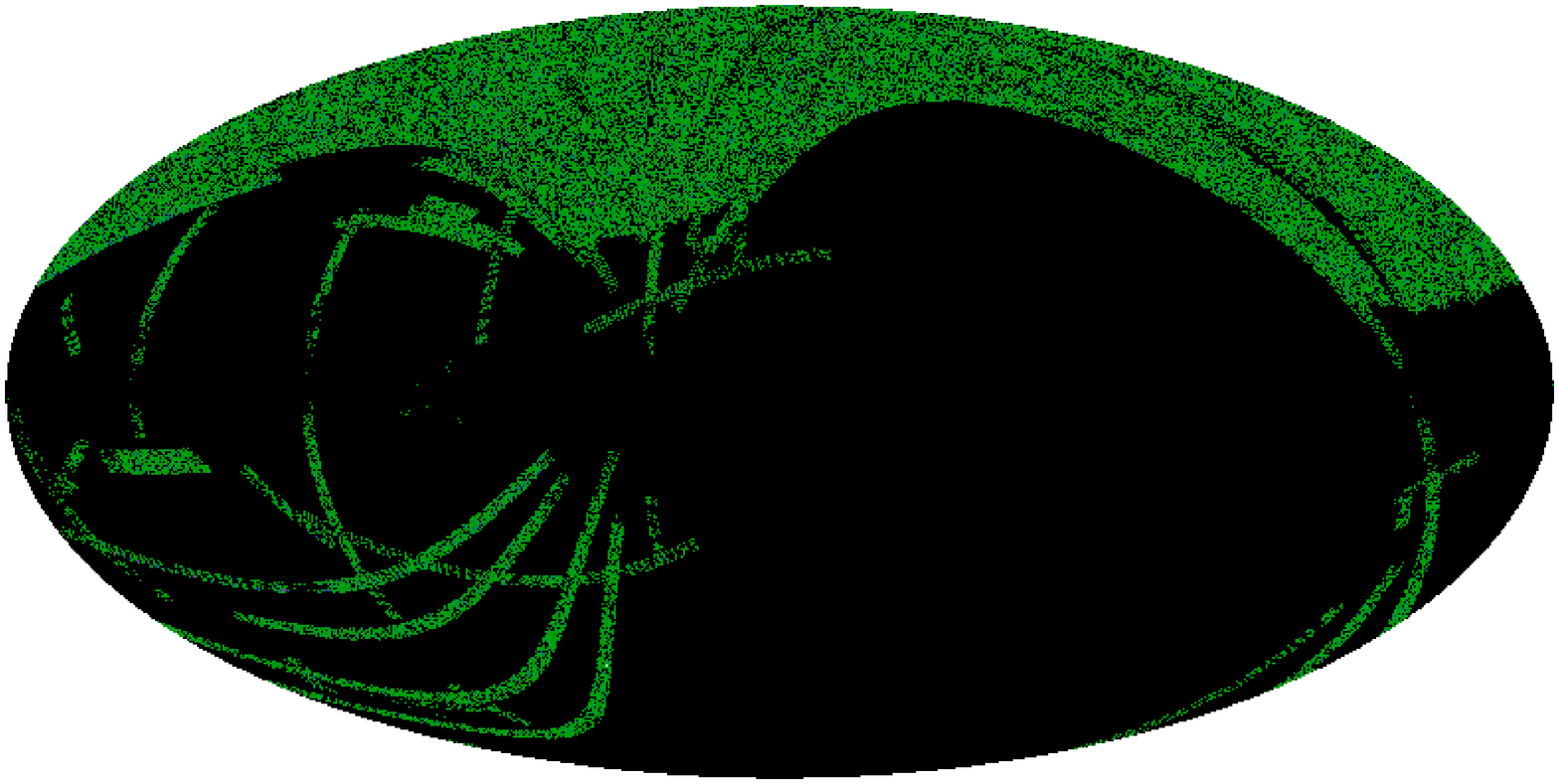}}
\resizebox{80mm}{!}{\includegraphics[angle=90.0]{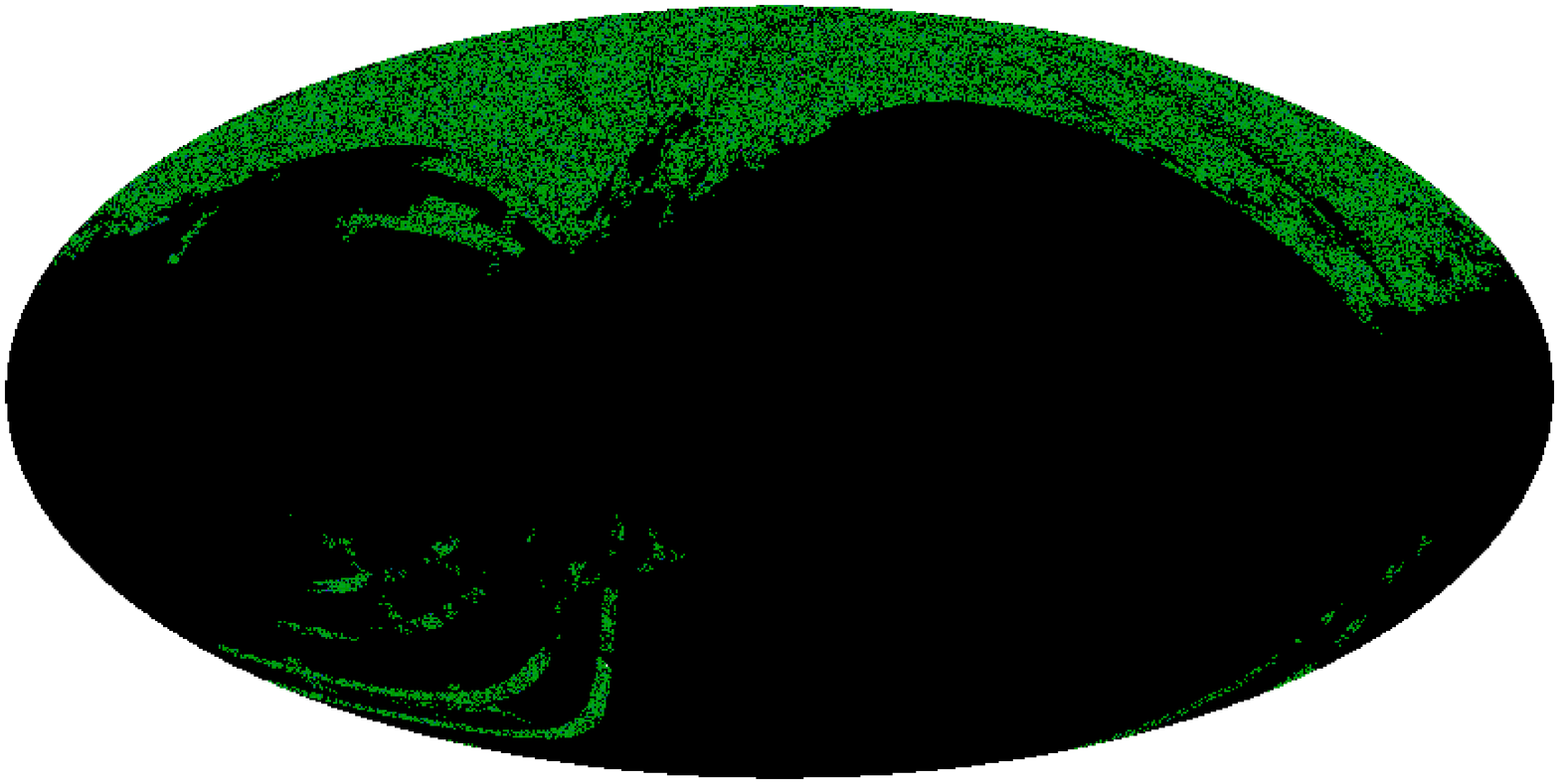}}\\
\resizebox{80mm}{!}{\includegraphics[angle=90.0]{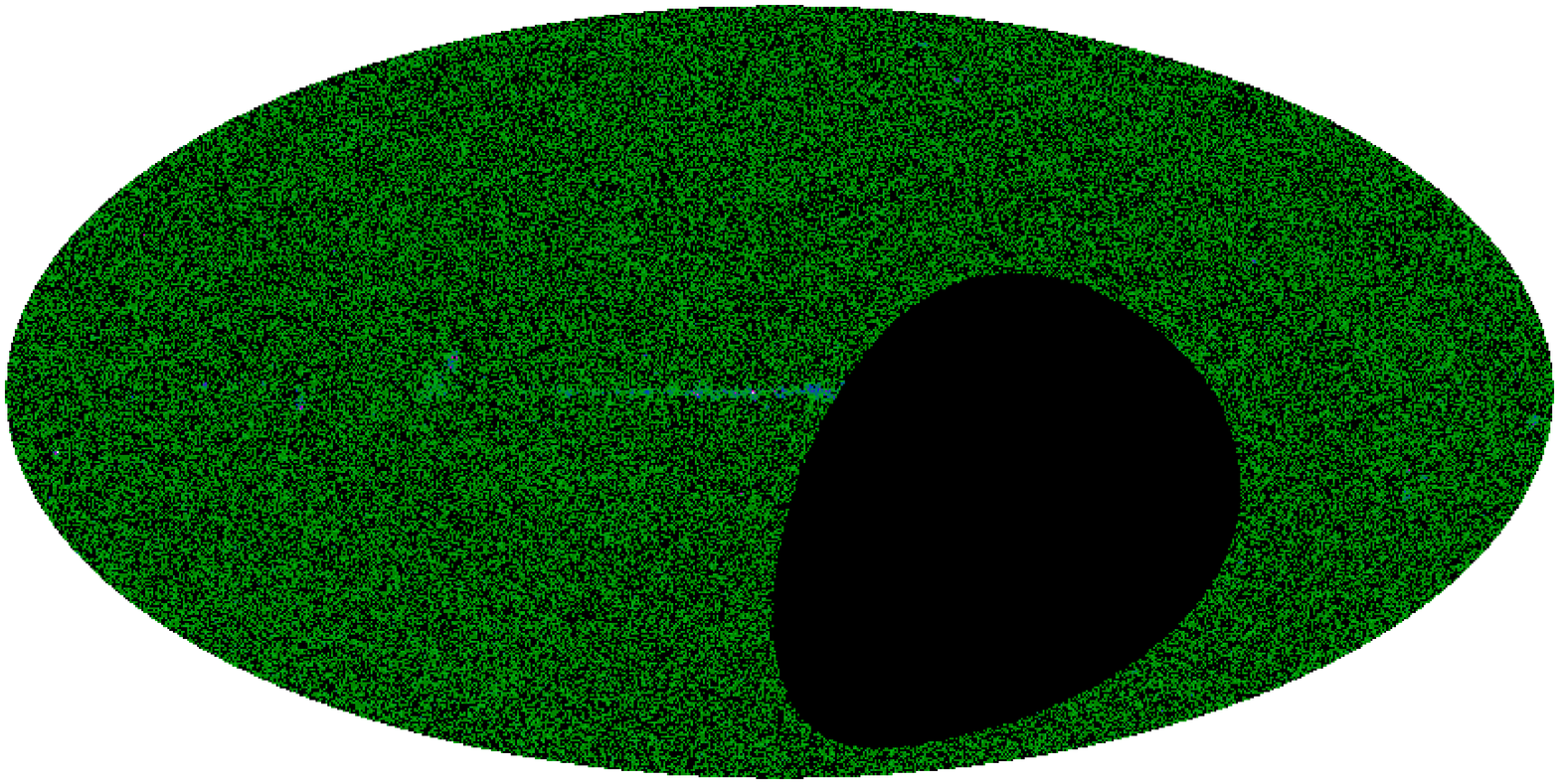}}
\resizebox{80mm}{!}{\includegraphics[angle=90.0]{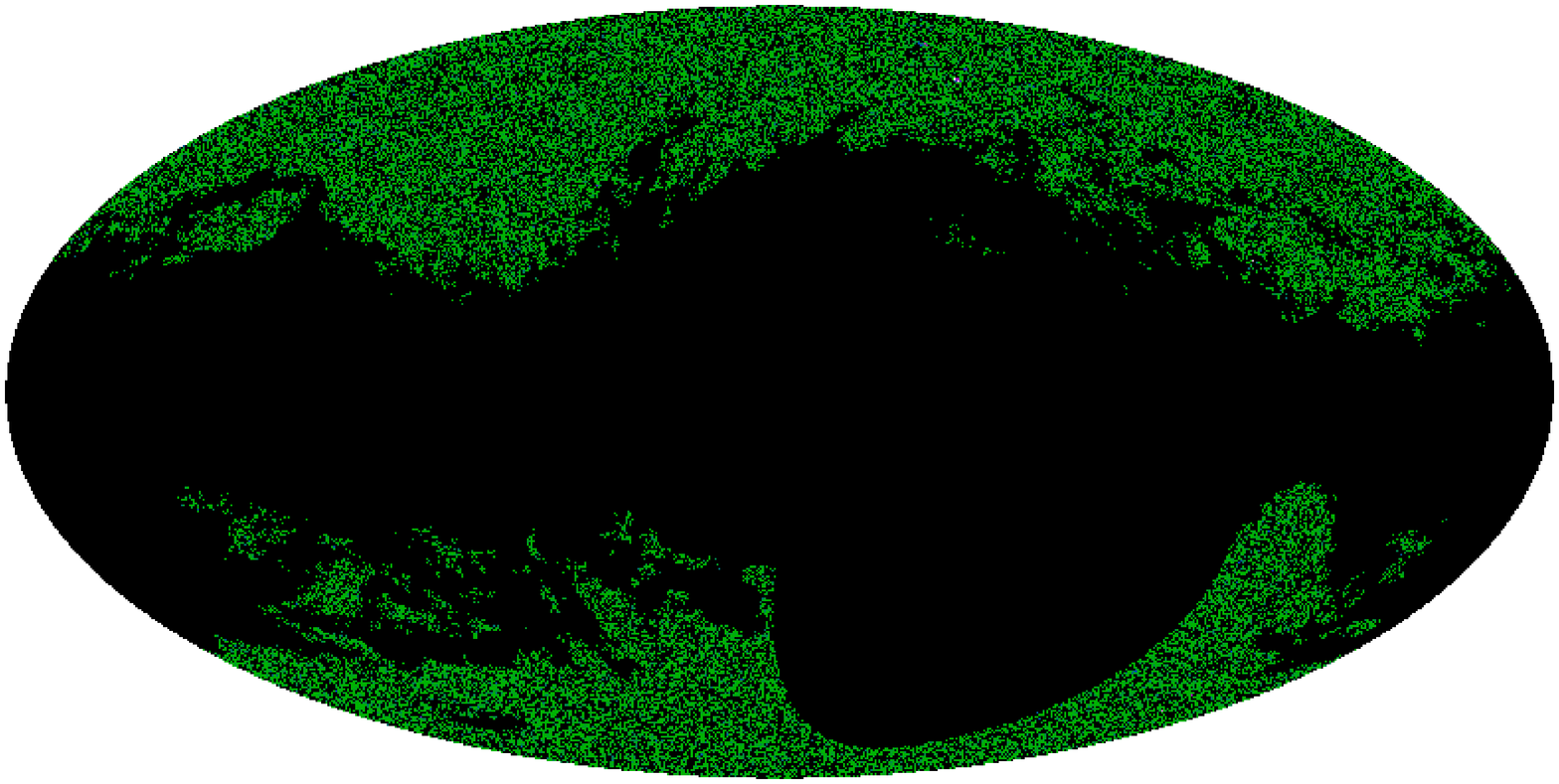}}\\
 \caption{Sky coverage of SDSS (top row) and NRAO VLA Sky Survey (NVSS) (bottom row) in Mollweide projection. The right column shows the same maps after applying the
radio (2 mJy) mask (SDSS only) and dust ($E(B-V) > 0.05$) mask (both). }
\end{center}
\label{coverage_fig}
  \end{figure*}

\section{Data Sets}

The analysis presented here uses four data sets: the Wilkinson Microwave Anisotropy Probe (WMAP) CMB temperature maps, the SDSS quasar and luminous red galaxy (LRG) catalogs, and the NRAO VLA Sky Survey (NVSS) radio catalog. Table 1 summarizes the sky areas, objects, and redshift ranges for these surveys. Figure 1 displays
the sky regions covered by the SDSS and NVSS catalogs, as well as the masks used in the analysis. 

\subsection{WMAP Temperature Maps}
The WMAP satellite has mapped the microwave sky in 5 frequency bands (Bennett et al.\ 2003) with an angular resolution ranging from 0.88 degrees to 0.22 degrees (corresponding to multipole moments ranging from $l=250$ to $l=900$. We have used the WMAP 5-year foreground-reduced CMB temperature maps for Q (41 GHz), V (61 GHz), and W (94 GHz) bands (Gold et al.\ 2009). The temperature $T_i$ assigned to pixel $i$ is the temperature difference from the mean map temperature, after dipole and foreground removal. These maps use the HEALPix scheme developed by Gorski et al.\ (2000). We have used the HEALpix resolution 9 maps for our analysis. This corresponds to a total of 3145728 pixels with each pixel being 47.2 square arc minutes in area. For each frequency band we use a weighted average of the maps from the individual differencing assemblies 
(Bennett et al. 2003). 

\subsection{SDSS Quasar Catalog}
SDSS has performed 5 band (ugriz) photometry (Fukugita et al.\ 1996) over roughly 10000 square degrees of sky area. We use the quasar catalog developed by Ho et al.\ (2008). The quasars are selected photometrically. A quasar candidate catalog of UV-excess objects ($(u-g) < 1.0$) with observed g magnitudes fainter than 14.5, extinction corrected g magnitudes brighter than 21.0, and u-band error less than 0.5 mag are generated (Ho et al.\ 2008). A catalog of DR3-UV-excess objects is also constructed. The candidate quasar catalog is then matched in color space with the DR3-QSO catalog (Richards et al.\ 2006) and the DR3-UV-excess catalog to generate the current QSO catalog. Three additional maps were used to clean the data set from stellar contamination and mask out regions affected by poor seeing: (i) a point spread function map; (ii) a stellar density map ($18.0 < r < 18.5$ stars smoothed with a 2-degree FWHM); and (iii) a stellar surface density map using red stars with $(g-r)$ greater than 1.4. These maps are described in detail in Ho et al.\ (2008). The data set here uses SDSS PSF magnitudes for all quasars.
 
\subsection{SDSS Luminous Red Galaxy Catalog}
Luminous red galaxies are useful mass tracers in the Universe. The LRGs tend to have old stellar population with a uniform spectral energy distribution. The characteristic 4000 \mbox{\AA} break makes LRGs ideal objects for photo-$z$ algorithms with redshift accuracies $\sigma_{z} \approx 0.03$ (Padmanabhan et al.\ 2005). The selection criterion, color and magnitude cuts for the LRG sample employed here are described in Padmanabhan et al.\ (2005). The LRGs are chosen to be galaxies that have colors consistent with an old stellar population, and absolute magnitudes above a certain threshold (Padmanabhan et al.\ 2005). The magnitude cuts are given as $r_{\rm petro} < (13.6 + 2.3(g-r) + 4(r-i-0.18))$, $r_{\rm petro} < 19.7$, $i < (18.3 + 2(r-i) - 0.25(g-r))$, and $i < 20$ where $r_{\rm petro}$ is the SDSS $r$-band Petrosian magnitude (Padmanabhan et al.\ 2005). The LRG catalog used for the current work is drawn from the SDSS sample prepared by Ho et al.\ (2008). The magnitudes we use for the LRGs are model magnitudes (as opposed to the PSF magnitudes used for quasars). 

\subsection{NVSS Catalog}
NVSS is a 1.4 GHz  continuum sky survey with the Very Large Array (VLA) (Condon et al.\ 1998). It covers the sky region that lies north of $\delta =-40^{\circ}$. The entire survey area is about 82\% of the celestial sphere. The survey catalog includes about $2 \times 10^{6}$ discrete objects with a full width half maximum of $45''$  and a nearly uniform sensitivity. The error in RA and dec varies from $< 1''$ for sources with fluxes above 15 mJy to about $7''$ for sources at the survey flux limit of 2 mJy (Condon et al.\ 1998). The median redshift for the NVSS objects is around $z=1$. 

\begin{table}
\begin{center}
\caption{Data Set Specifications}
\begin{tabular}{c|c|c|c}
\hline
\hline
\multicolumn{1}{c|}{Catalog}&
\multicolumn{1}{c|}{Redshift}&
\multicolumn{1}{c|}{Area}&
\multicolumn{1}{c}{Objects}\\
\multicolumn{1}{c|}{}&
\multicolumn{1}{c|}{}&
\multicolumn{1}{c|}{(deg$^2$)}&
\multicolumn{1}{c}{}\\
\hline
 QSO catalog      & 0.08-2.82 & 6039  & 586435 \\ 
 LRG catalog    & 0.4-0.6  & 6641  & 911686 \\
 NVSS catalog   & 0-3.0      & 27361 & 1104983\\ 
 \hline 
\end{tabular}
\end{center}
\end{table}

\section{The Cross-Correlation and Systematics}

We have cross-correlated the SDSS QSO catalog with the WMAP sky maps to estimate the average
amplitude of the SZ distortion associated with a known QSO.
The estimated amplitude of the quasar signal in a given WMAP band is simply 
\begin{equation}
\Delta T = \left(\sum_{i}N_{i}T_{i}\right) \left(\sum_{i}N_{i}\right)^{-1},
\label{TQdef}
\end{equation}
where $N_{i}$ is the number of quasars in the ith WMAP pixel and $T_{i}$ is the effective thermodynamic temperature of the pixel in the chosen frequency band.
This expression just gives the average temperature distortion associated with each individual
QSO.  The theoretical SZ signal from a single QSO is expected to be about 1-10 $\mu$K for an arcminute beam depending on the mass of the black hole (Scannapieco, Thacker, \& Couchman 2008; Chatterjee et al.\ 2008). This when translated to WMAP beams gives a signal of roughly 10-100 $\mu$Jy at WMAP frequencies (Refreiger, Spergel, \& Herbig 2000). At WMAP frequencies, the SZ effect reduces the temperature in a pixel, so the average temperature
in Eq.\ 1 will be negative if dominated by the SZ effect. Galactic foregrounds and other systematic emission (described later) will positively bias the signal and lead to an underestimate of the
SZ temperature decrement. For comparison, we also cross-correlate the SDSS LRG catalog with the WMAP
sky maps, which serves as a rough measure of the galaxy cluster SZ signal to the extent that the LRG sample
predominantly traces clusters. 


Signals contributing to the sum in Eq.\ 1 are affected by a number of possible
systematic effects, which we attempt to minimize. Sky regions with higher dust extinction will lead to selection biases in the QSO and LRG samples. Also, dust emission in these bands will contaminate the WMAP signal. 
We employ the dust prescription described in Ho et al.\ (2008). Using the Schlegel, Finkbeiner, \& Davis (1998) extinction map, we construct E(B-V) masks excluding pixels with E(B-V) above some threshold value (see Table 3 for the dust mask thresholds). We apply the E(B-V) mask to both our SDSS and NVSS samples. Figure 1 shows the SDSS and NVSS sample spaces after applying the dust masks. 

Another source of contamination in the signal will come from radio emission from radio-loud quasars. The spectral index of these radio-loud sources is commonly around $0.7$ (Carlstrom, Holder \& Reese 2002),
so radio contamination at low frequencies is an important issue. 
We have matched the objects in the SDSS quasar catalog with the NVSS radio catalogs to find radio counterparts of the Sloan quasars. This was accomplished by (note that WMAP sky pixels are 7 arcminutes) 
choosing radio sources which are within an angular distance of 45 arcseconds, the beam size of the radio survey (although the pointing error
is substantially less). 
We then mask WMAP pixels that contain selected radio sources with fluxes above 2 mJy at 1.5 GHz. The right column of Figure 1 shows the Mollweide projections of our data sets after applying the dust mask and the radio mask (for the SDSS samples only).  

We also want to minimize the noise arising from random associations of QSO positions with
temperature distortions in the primary CMB anisotropies.
Following the WMAP team's spatial Weiner filter analysis when obtaining point source amplitudes from the temperature maps (Hinshaw et al.\ 2007), we use a similar high-pass spatial filter to cut off power at large angular scales where primary CMB fluctuations dominate the signal. Microwave temperature maps are
conventionally expressed as spherical harmonic coefficients, 
$T(\theta,\phi)=\sum_{lm} a_{lm}Y_{lm}(\theta,\phi)$. A filtered map is reconstructed from the coefficients
$a'_{lm} = {\cal F}_l a_{lm}$, with the Wiener filter (Tegmark \& de Oliveira-Costa 1998; also see Chatterjee 2009 for detailed implementation of the filter)
\begin{equation}
{\cal F}_l = \left(\frac{4\pi W_l}{c_l W_l^2 + n_l}\right) \left(\sum_l\frac{(2l+1)W_l^2}{c_l W_l^2 + n_l}\right)^{-1},
\label{filter}
\end{equation}
where $c_l$ are the multipole moments for the primary CMB power spectrum, $W_l$ are the
moments of the angular WMAP beam profile, and $n_l$ give the detector noise at each multipole moment. 
We computed the $c_{l}$ values from the best-fit WMAP cosmological model
using the CAMB code (Lewis, Challinor \& Lasenby 2000) (although the filter is relatively insensitive to the precise set of multipole moments used), the detector noise $n_l$ from the
noise-per-differencing-assembly values for WMAP (WMAP explanatory supplement: Limon et al.\ 2009), and the WMAP beam moments
from Page et al.\ (2003). The maximum $l$ values that we have used for constructing the filters for different frequencies depend on the angular resolutions of the corresponding bands: $l=1000$ (Q-band), 
$l=1500$ (V-band), and $l=2000$ (W-band).  The filter suppresses power at scales which are dominated
by the primary CMB temperature fluctuations or detector noise; the precise form of the filter in Eq.\ 2
gives a better signal-to-noise estimate of the amplitude of a point source superimposed on the primary CMB fluctuations and observed with the WMAP beam and detector noise.  At WMAP angular 
resolutions, the SZ signal from an individual QSO or galaxy cluster can be treated as a point source.

\section{Results}

\begin{table}
\begin{center}
\caption{Fits for the Spectrum}
\begin{tabular}{c|c|c|c|c}
\hline
\hline
\multicolumn{1}{c|}{Sample}&
\multicolumn{1}{c|}{Model}&
\multicolumn{1}{c|}{Fits for $y$}&
\multicolumn{1}{c|}{$\chi^{2}$}&
\multicolumn{1}{c}{BIC}\\
\multicolumn{1}{c|}{}&
\multicolumn{1}{c|}{}&
\multicolumn{1}{c|}{$10^{-7}$}&
\multicolumn{1}{c|}{}&
\multicolumn{1}{c}{}\\
\hline
 QSO &SZ     &$y=1.6 \pm 1.6$&$2.58$ &$5.28$\\
\hline 
 QSO& SZ+Dust&$y=7.0 \pm 3.4$&$2.45$&$4.65$\\
\hline
LRG & SZ&$y=0.02 \pm 0.7$&$2.75$&$6.59$\\    
\hline
LRG & SZ+Dust&$y=5.3 \pm 2.45$&$2.78$&$4.98$\\      
\hline 
\end{tabular}
\end{center}
\end{table} 

\begin{figure}
\begin{center}
\begin{tabular}{cc}
\resizebox{80mm}{!}{\includegraphics{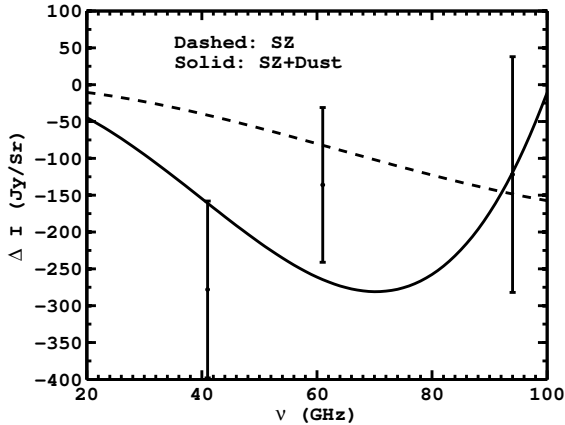}}\\
\end{tabular}
   \caption{The cross-correlation spectrum of the QSOs. The dashed  line is the best-fit SZ spectrum,
   while the solid line is the best two-component fit using both SZ and dust. The best-fit values for the $y$ parameter and the corresponding reduced $\chi^{2}$ values are listed in Table 2.}
\end{center} 
\label{spectrum_fig_qso}
  \end{figure}

Figure 2 shows the average radiation intensity difference per QSO in three frequency bands, obtained from Eq.\ 1. For clarity across bands, the thermodynamic effective temperature in each band is converted to intensity. Figure 3 shows the same signal, except for
the cross-correlation with LRG's instead of QSO's. The dashed  lines shows the best-fit SZ distortion with the usual spectrum 
\begin{equation}
\frac{\Delta I}{I_{0}} = \frac{x^{4}e^{x}}{(e^{x}-1)^{2}}\left[x\coth(x/2)-4\right]y,
\label{SZspectrum}
\end{equation} 
where $x=h\nu/(K_{B}T_{0})$. The solid lines are a two-component fit incorporating both SZ and dust emission
with a spectrum $I(x)= Ax^{4}$, corresponding to an effective dust temperature power law index of 2 (Gold et al.\ 2009). 
The best-fit $y$ values for all cases 
are shown in Table 2. The errors on the fit parameters are generated in the following way. We resimulate the data by drawing an error from a Gaussian distribution with root-mean-squared values given by the standard errors in the estimated signal for each band calculated from the variance in $\Delta T$ and number of pixels used for averaging. We then perform the fits on the mock data and determine the standard deviation of the fit parameters among all mock datasets. The value of this standard deviation is provided as the estimate of error on the parameter. The values for the reduced $\chi^{2}$ and the Bayesian Information Criterion (BIC $ = \chi^{2}+k\log(n)$; k: degrees of freedom; n: number of data points; Liddle, Mukherjee, \& Parkinson 2006) for the various cases are also listed in Table 2. The information criterion tests whether the improvement in the fits is worth the extra parameters. The two-component fit gives a lower $\chi^2$ value;  the improvement to the BIC indicates that the improvement from adding an additional parameter is enough to outweigh the extra model freedom from the additional parameter.
This conclusion should perhaps be taken with a grain of salt since only three data points with large errors are being fit.  Note also that the two-parameter fit gives different values of $y$: the larger
SZ component can better fit the suppressed point at 41 GHz, and the resulting lower SZ signal at 94 GHz
can then be compensated by the increased dust contribution. The overall spectra for the two possibilities
are significantly different, and measurements with smaller errors and data points at higher frequencies
can distinguish between the two cases. 

To investigate the significance of the observed anticorrelation, we randomize the positions of the quasars and recalculate the values of $\Delta I$ for a thousand realizations. We find that in $0.1\%$ (Q band), $1.3\%$ (V band), and $6\%$ (W band) of all realizations, the correlation amplitude is more negative than 
in the real sample. The E(B-V) $\geq 0.05$ mask has been used in this case. We are thus confident that we are observing a real average CMB temperature decrement associated with QSO's. We change the thresholds of the dust masks in calculating the correlation for the randomized case. The results do not change with the change in the thresholds of the dust masks.  

\begin{table*}
\begin{center}
\caption{Effect of Varying the Dusk Mask}
\begin{tabular}{c|c|c|c}
\hline
\hline
\multicolumn{1}{c|}{Band}&
\multicolumn{1}{c|}{E(B-V) $\geq 0.01$}&
\multicolumn{1}{c|}{E(B-V) $\geq 0.05$}&
\multicolumn{1}{c}{E(B-V) $\geq 0.10$}\\
\hline
 Q      &$-739 \pm 468$ Jy/sr & $-278 \pm 120$ Jy/sr & $-229 \pm 114$ Jy/sr \\
 V      &$-1004 \pm 414$ Jy/sr & $-136 \pm 105$ Jy/sr & $-124 \pm 97$ Jy/sr \\
 W      &$-30 \pm 616$ Jy/sr & $-122 \pm 160$ Jy/sr & $-46 \pm 147$ Jy/sr \\
\hline 
\end{tabular}
\end{center}
\end{table*}
\begin{table*}
\begin{center}
\caption{Effect of Varying the Dusk Mask with Randomized QSO and LRG positions}
\begin{tabular}{c|c|c|c}
\hline
\hline
\multicolumn{1}{c|}{Band}&
\multicolumn{1}{c|}{E(B-V) $\geq 0.01$}&
\multicolumn{1}{c|}{E(B-V) $\geq 0.05$}&
\multicolumn{1}{c}{E(B-V) $\geq 0.10$}\\
\hline
 Q (QSO) &$0.12$ Jy/sr & $0.50$ Jy/sr & $0.84$ Jy/sr \\
 V (QSO) &$0.19$ Jy/sr & $0.85$ Jy/sr & $1.23$ Jy/sr \\
 W (QS0) &$1.6$ Jy/sr & $8.2$ Jy/sr & $10.9$ Jy/sr \\
 Q (LRG) &$0.01$ Jy/sr & $0.10$ Jy/sr & $0.01$ Jy/sr \\
 V (LRG) &$0.0$ Jy/sr & $0.12$ Jy/sr & $0.81$ Jy/sr \\
 W (LRG) &$5.8$ Jy/sr & $10.12$ Jy/sr & $12.1$ Jy/sr \\
\hline 
\end{tabular}
\end{center}
\end{table*}

To study the possible impact of foreground emission or absorption, we investigate the change in the cross-correlation amplitude when we change thresholds used for each mask. Table 3 shows the effect on the cross-correlation amplitude of changing the threshold of our dust mask. The signal in the Q and V bands
varies significantly with the thresholds of the dust mask; we note that the cross-correlation amplitude increases in these bands (that is, it becomes more negative) when we use more restrictive masks. The signal in the W band is always consistent with noise. The fact that the cross-correlation signal does not converge with
increasingly stringent dust map masks indicates that the observed signal is due to a combination of SZ
and dust components. We speculate that some of the actual SZ decrement is filled in by dust emission in regions with higher dust. Also since dust will not be correlated with the quasar positions, the mean dust signal contributions to the QSO and LRG sample should be comparable. To check this we examined the trend of the signal with the threshold of the dust mask using the randomized sample for both LRGs and QSOs. The results are shown in Table 4.0. The values for the LRG and QSOs are rougly consistent and also we see the correlation value to be high positive (compared to the other bands) in W band. 
This likely means that the $y$-parameters derived
here for the two-component fit including dust are closer to the real actual $y$-parameters.
Our cross-correlation signals are unchanged when the radio mask threshold is changed
from 2 mJy to 4 mJy, so the signal likely has no significant radio contamination. This is further born out because
the signal is unchanged if we increase the coverage of the mask to include all WMAP pixels
within 20 arcminutes of any radio source.    

One of the major possible contributors to the cross-correlation signal is an SZ distortion from galaxy clusters associated with the QSOs. A number of authors have previously reported evidence of SZ distortion in WMAP temperature maps (e.g; Bennett et al.\ 2003; Myers et al.\ 2004; Hernandez-Monteagudo et al.\ 2006; Spergel et al.\ 2007; Diego \& Partridge 2010), most likely from unresolved galaxy clusters. To estimate the effect from galaxy clusters, we cross-correlate SDSS LRGs, which serve as a tracer of galaxy clusters, with the same
filtered WMAP maps and the same masks used for the QSOs. Errors are also calculated in the same way; the
signal has about the same statistical significance as the QSO signal.
The cross-correlation spectrum is shown in Fig.\ 3; the amplitude and shape of the distortion is very similar to the QSO case shown in Fig.\ 2. The similarity in the cross-correlation spectrum suggests that both the signals might be coming from the same SZ source population. 

Chatterjee et al.\ (2008) shows that the SZ signal from QSOs will be dependent on redshift due to the time evolution of quasar feedback. We perform this test by binning the quasars in two redshift slices ($z \leq 1.5$ and $z \geq 1.5$). The results do not show any significant evolution with redshift. 
\begin{figure}
\begin{center}
\begin{tabular}{cc}
\resizebox{80mm}{!}{\includegraphics{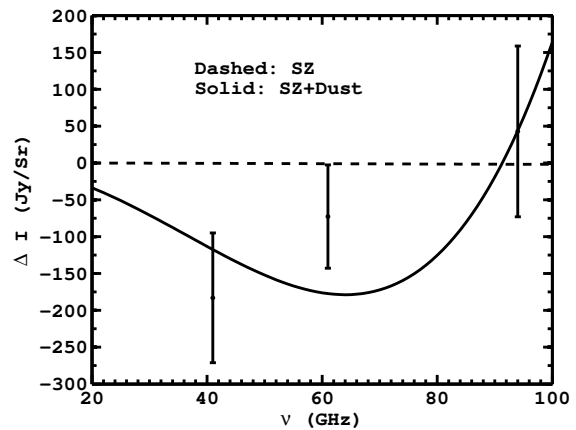}}\\
\end{tabular}
   \caption{The cross-correlation spectrum of the LRGs.}
\end{center} 
\label{spectrum_fig_lrg}
  \end{figure}
       
\section{Discussion and Prospects}
 
The SZ effect serves as a viable probe for detecting hot gas in AGN environments, though the signal
is small. We have presented a first step toward detecting this signal by cross-correlating optical quasar catalogs with microwave maps. At frequencies lower than the null frequency, the SZ signal is a decrement which makes it unique among other potential signals at those frequencies. However, oversubtraction of foregrounds can also lead to decrements. Wright et al.\ (2009), Hinshaw et al.\ (2007), and Chen \& Wright (2008) report detection of negative fluxes associated with point sources in the WMAP 3-year and 5-year point source catalogs. In the present work we find a deficit in microwave flux associated with both QSO's
and LRG's from SDSS, which suggests SZ, but the fits to the observed spectrum are not sufficiently constraining to establish this conclusion definitively. 

Any SZ signal is likely some combination of the signal associated with gas heated by QSO's and
hot gas in galaxy clusters. 
The observed correlation of flux decrement with the LRGs is similar to that for QSOs. This suggests that the sources of the signals might be similar. LRGs are tracers of galaxy clusters, and hence the SZ effect that we observe could be originating from galaxy clusters associated with QSOs. Diego \& Partridge (2010) have
analyzed WMAP data by stacking regions around known X-ray clusters, finding an SZ signal from galaxy clusters in the range of 10 to 20 $\mu$K, corresponding to tens of mJy.  The mean fluxes we get are of the order of 1 mJy, while the theoretical mean SZ signal from quasars is in the range 10 to 100 $\mu$Jy, depending on the mass of the black hole (Chatterjee \& Kosowsky 2007; Scannapieco, Thacker, \& Couchman 2008). Our signal is at least an order of magnitude higher than the signal that we expect from quasars and an order of magnitude less than what is expected from massive galaxy clusters. 

The median redshift of the quasars in our sample is $1.3-1.4$; quasars at these redshifts tend to reside in halos with masses around a few 
$10^{12}\, M_\odot$ (e.g., Coil et al.\ 2007) and not in cluster mass haloes. However, since the cluster SZ signal is so much larger than the QSO SZ signal, our results can be explained if only a small fraction of QSO's reside
in galaxy clusters. If 10\% of QSOs reside in clusters with $10^{14}$ to $10^{15}\, M_{\odot}$ halos, they would
provide the observed cross-correlation signal. We made an estimate of the cluster signal from a typical halo mass for QSOs $10^{12} M_{\odot}$ and a typical cluster mass $10^{14} M_{\odot}$ by estimating the angular cross correlation function $\Pi_{qso,cluster}(\theta)$, which will be equal to bias (QSOs) x bias ($10^{14}$ solar mass halos) x mean dark matter angular correlation function at $z \approx 1$. To get an estimate of the SZ contribution of the cluster signal we need to integrate the correlation function over the beam size which in Jy/Sr will be,$\left(\pi \theta_{beam}^{2}\right)^{-1}\int_{0}^{\theta_{beam}}d\theta 2\pi \theta \Pi_{qso,cluster}(\theta)\times$(mean cluster SZ decrement)$\times$(angular number density of clusters). Assuming a beam size of 20 arcminutes for WMAP, $b_{QSO} =2.31$ (Shen et al.\ 2009), $b_{cluster} =3.3$ (Estrada et al.\ 2009), a cluster density of 10 per square degrees (Carlstrom, Holder \& Reese 2002), the angular correlation function estimate at redshift 1.0 (Abell et al.\ 2009), and a typical SZ signal from a cluster to be $\sim 20 mJy$ we get an estimate of the cross-correlation to be $\sim 65 Jy/Sr$. This number is roughly comparable to our estimates of the cross-correlation. However this estimate is slightly lower than our detected signal which might imply that some of the power in the signal could be still related to backgrounds. An alternate possibility is that the SZ signal from quasar feedback is higher (Natarajan \& Sigurdsson 1999) than those obtained from current simulations  (Chatterjee et al.\ 2008; Scannapieco, Thacker \& Couchman 2008). 

Higher-resolution CMB maps can distinguish between QSO and galaxy cluster SZ signals by simply
resolving the cluster signals (to all redshifts) and removing these pixels from any cross-correlation.
Experiments which may improve on the current cross-correlation analysis include the Planck
satellite, with angular resolution down to a few arcminutes in its higher frequency bands, and
the ground-based arcminute resolution experiments like ACT, SPT, and APEX. 
The SZ effect is subdominant at angular scales probed by WMAP (e.g., Komatsu \& Seljak 2002; Bennett et al.\ 2003), which will not be the case at higher resolution. The mean QSO SZ amplitude is predicted to be on
the order of 1 $\mu$K at 150 GHz for an arcminute size beam (Scannapieco, Thacker \& Couchman 2008), with substantial
modeling uncertainty. Data from SDSS reveal around 30 photometrically detected quasars per square degree. 400 square degrees of sky coverage can provide cross-correlation with around 10000 QSO's. 
If a ground-based experiment can attain a noise of 10 $\mu$K per pixel over this sky area,
this will give roughly a $10\sigma$ detection of the QSO cross-correlation signal and a clear separation of the
cluster and QSO SZ signals.  Planck will have 7-arcminute resolution at 143 GHz; this implies a mean QSO SZ decrement of around 40 nK per pixel. The noise per pixel after 14 months is projected to be
around 6 $\mu$K. With around $5\times 10^5$ quasars over the full sky, the signal-to-noise of
the cross-correlation will be $5\sigma$.  Note that both of these estimates are for single frequency bands
at the maximum decrement. Planck and ACT in particular are both mapping at frequencies above
the null with higher angular resolution, giving stronger statistical constraints but more difficult
systematic errors due to possible confusion between the SZ increment and foreground emission. 
Planck's many frequency bands may allow efficient separation of SZ from foreground through
spectrum information.       

We have demonstrated a statistically significant correlation of a temperature decrement in WMAP
sky pixels with the number density of both quasars and luminous red galaxies drawn from the Sloan Digital
Sky Survey. The inferred QSO $y$-distortion for a two-component fit including both dust and SZ gives
$y=(7.0\pm 3.4)\times 10^{-7}$. The signal appears to be partly
correlated with local dust emission. Higher signal-to-noise measurements are required to
separate the dust and SZ contributions, while higher spatial resolution will separate QSO and
galaxy cluster SZ signals. Both prospects are likely in the near term with the Planck satellite
and dedicated ground-based SZ measurements. We anticipate that SZ measurements will
soon make a valuable contribution to characterizing the energy feedback from quasars, to
complement current data from the X-ray and radio bands.


\begin{acknowledgments}
SC would like to thank David Spergel for thoughtful suggestions throughout this work, 
Neelima Sehgal and Zheng Zheng for helpful discussions, 
Grant Wilson for insights about the observed spectrum, and Valery Rashkov for providing his undergraduate thesis from Princeton University which contains a useful derivation of the Weiner filter. SC would also like to thank the Department of Astrophysical Sciences at Princeton University for hosting her visits to Princeton and Adam Solomon for helping with some of the analysis codes. We would also like to thank the referee for providing valuable feedback of our work which helped in significant improvement of the draft. We acknowledge Craig Markwardt for usage of the MPFIT package, and the Legacy Archive for Microwave Background Data Analysis (LAMBDA) for providing the data products from the WMAP Science Team. This work was supported at the University of Pittsburgh by the National Science Foundation through grant AST-0408698 to the ACT project, and by grants AST-0546035 and AST-0807790. SC was partly funded by the Zaccheus Daniel Fellowship and the Andrew Mellon Fellowship at the University of Pittsburgh. Funding for the SDSS and SDSS-II has been provided by the Alfred P.\ Sloan Foundation, the Participating Institutions, the National Science Foundation, 
the U.S.\ Department of Energy, the National Aeronautics and Space Administration, the Japanese Monbukagakusho, the Max Planck Society, and the Higher 
Education Funding Council for England. The SDSS Web Site is http://www.sdss.org/.

The SDSS is managed by the Astrophysical Research Consortium for the Participating Institutions. The Participating Institutions are the American Museum 
of Natural History, Astrophysical Institute Potsdam, University of Basel, University of Cambridge, Case Western Reserve University, University of 
Chicago, Drexel University, Fermilab, the Institute for Advanced Study, the Japan Participation Group, Johns Hopkins University, the Joint Institute 
for Nuclear Astrophysics, the Kavli Institute for Particle Astrophysics and Cosmology, the Korean Scientist Group, the Chinese Academy of Sciences 
(LAMOST), Los Alamos National Laboratory, the Max-Planck-Institute for Astronomy (MPIA), the Max-Planck-Institute for Astrophysics (MPA), New Mexico 
State University, Ohio State University, University of Pittsburgh, University of Portsmouth, Princeton University, the United States Naval Observatory, 
and the University of Washington.
\end{acknowledgments} 
\pagebreak
\section{REFERENCES} 
~~\\
Abell, P., et al.\, 2009, LSST Science book, galaxy chapter, arXiv 0912.0201L\\ 
Afshordi, N., Loh, Y., \& Strauss, M., 2004, PhRvD, 69, 083524\\
Aghanim, N., Balland, C., \& Silk, J., 2000, A\&A, 357, 1\\
Babich, D. \& Loeb, A. 2007, \mnras, 374, L24\\
Bennett, C. L., et al., 2003, ApJS, 148, 1\\
Birzan, L., Rafferty, D. A., McNamara, B. R., Wise, M. W., \& Nulsen, P. E. J., 2004, ApJ, 607, 800\\
Carlstrom, J., E., Holder, G., P., \& Reese, E. D., 2002, ARA\&A, 40, 643\\
Chatterjee, S., \& Kosowsky, A., 2007, ApJL, 661, L113 \\
Chatterjee, S., Di Matteo, T., Kosowsky, A., \& Pelupessey, I., 2008 \mnras, 390, 535\\
Chatterjee, S., 2009, etd-07072009-133501, (University of Pittsburgh Press), \\
http://etd.library.pitt.edu/ETD/available/etd-07072009-133501/\\
Chen, X., Wright, E. L., 2008, ApJ, 681, 747\\
Cheng, L. M., Wu, X. P., \& Cooray, A., 2004, A\&A, 413, 65\\
Coil, A. L., Hennawi, J. F., Newman, J. A., Cooper, M. C., \& Davis, M., 2007, \apj, 654, 115\\
Condon, J. J., et al., 1998, \apj, 115, 1693\\
Croft, R. A. C., Banday, A. J., \& Hernquist, L., 2006, \mnras, 369, 1090\\
de Zotti, G., Burigana, C., Cavaliere, A., Danese, L., Granato, G. L., Lapi, A., Platania, P., \& Silva, L., 2004, AIPC, 703, 375D\\
Diego, J. M., Silk, J., \& Sliwa, W., 2003, \mnras, 346, 940\\
Diego, J. M., \& Partridge, B, 2010, \mnras, 402, 1179\\
Dunn, R. J. H., Fabian, A. C., \& Taylor, G. B., MNRAS, 364, 1343\\
Dunn, R. J. H., \& Fabian, A. C., MNRAS, 2006, 373, 959\\
Estrada, J., Sefusatti, E., \& Frieman, J. A., 2009, \apj, 692, 265\\
Fosalba, P., Gaztanaga, E., Castander, F. J., 2003, \apj, 597, L89\\
Fukugita, M., Ichikawa, T., Gunn, J. E., Doi, M., Shimasaku, K., \& Schneider, D. P., 1996, 111, 1748\\
Giannantonio, T., Scranton, R., Crittenden, R. G. Nicol, R. C., Boughn, S. P., Myers, A. D., \& Richards, G. T., 2008, PhRvD, 77, 123520\\
Gold, B., et al., 2009, ApJs, 180, 265\\
Gorski, K. M., et al., 2005, \apj, 622, 759\\
Hernandez-Monteagudo, C., Macias-Perez, J. F., Tristram, M., \& Desert, F., X., 2006, A\&A, 449, 41\\
Hincks, A. D., et al.\, 2009, arxiv: 0907.0461\\
Hinshaw, G., et al.\, 2007, ApJS, 170, 288\\
Hirata, C. M., Padmanabhan, N., Seljak, U., Schlegel, D., \& Brinkmann, J., 2004, PhRvD, 70, 103501\\
Ho, S., Hirata, C. M., Padmanabhan, N., Seljak, U., \& Bahcall, N., 2008, PhRvD, 78, 043519 \\
Ho, S., Dedeo, S., Spergel, D., 2009, arxiv:0903.2845\\
Iliev, I.~T., Pen, U.-L., Bond, J.~R., Mellema, G., \& Shapiro, P.~R. 2006, New Astron.\ Rev.\ 50, 909.\\
Kashlinsky, A., Atrio-Barandela, F., Kocevski, D., \& Ebeling, D., 2009, \apj, 691, 1479\\
Komatsu, E., \& Seljak, U., 2002, \mnras, 336, 1256\\
Lapi, A., Cavaliere, A., \& De Zotti, G., 2003, ApJL, 597, L93\\
Liddle, A., Mukherjee, P., \& Parkinson, D., 2006, arxiv: astro-ph/0608184\\
Lewis, A., Challinor, A., Lasenby, A., 2000, \apj, 538, 473\\
Markwardt, C. B. 2008, in proc. ASP Conference Series (Astronomical Society of the Pacific: San Francisco) Arxiv preprint: arXiv:0902.2850v1\\
Majumdar, S. Nath, B., \& Chiba, M. 2001, \mnras, 324, 537\\
Massardi, M., Lapi, A., de Zotti, G., Ekers, R. D., \& Danese, L., 2008, \mnras, 384, 701\\
McNamara, B. R., Nulsen, P. E. J., Wise, M. W., Rafferty, D. A., Carilli, C., Sarazin, C. l., \& Blanton, E. L., 2005, Nature, 433, 45\\
McQuinn, M., Furlanetto, S.~R., Hernquist, L., Zahn, O., \& Zaldarriaga, M. 2005, \apj, 630, 643\\
Moodley, K., Warne, R., Goheer, N., \& Trac, H., 2008, arxiv:0809.5172\\
Myers, A. D., Shanks, T., Outram, P. J., Frith, W. J., \& Wolfendale, A. W., 2004, \mnras, 347, L67\\
Natarajan, P., \& Sigurdsson, S., 1999, \mnras, 302, 288\\
Oh, S.~P., Cooray, A., \& Kamionkowski, M. 2003, \mnras, 342, 20\\
Padmanabhan, N., Hirata, C. M., Seljak, U., Schlegel, D. J., Brinkmann, J., Schneider, D. P., 2005, PhRvD, 72, 043525\\
Page, L., et al. 2003, ApJS, 148, 39\\
Peebles, P. J. E., 1980, Princeton University Press\\
Peiris, H. V., \& Spergel, D.~N., 2000, \apj, 540, 605\\
Platania, P., Burigana, C., De Zotti, G., Lazzaro, E., \& Bersanelli, M., 2002, \mnras, 337, 242\\
Rashkov, V., 2009, Princeton University undergraduate thesis (unpublished). \\
Richards, G.~T.\ et al.\ 2006, \aj, 131, 2766\\
Refregier,A., Spergel, D. N., \& Herbig, T., 2000, ApJ, 531, 31\\
Rosa-Gonzalez, D., Terlvich, R., Terlvich, E., Friaca, A., \& Gaztanaga, E., 2004, \mnras, 348, 669\\
Roychowdhury, S., Ruszkowski, M., \& Nath, B.~B. 2005, \apj, 634, 90\\
Sanderson, A. J. R., Ponman, T. J., \& O'Sullivan, E., 2006, \mnras, 372, 1496\\
Scannapieco, E., Thacker, R. J., Couchman, H. M. P., 2008, \apj, 678, 674 \\
Schawinski, K., Tomas, D., Sarzi, M., Maraston, C., Kaviraj, S., Joo, S., Yi, S., Silk, J., 2007, \mnras, 382, 1415\\
Schlegel, D. J., Finkbeiner, D. P., \& Davis, M., 1998, \apj, 500, 525\\
Shen, Y., et al.\, 2009, \apj, 697, 1656S\\
Smith, K. M., Zahn, O., \& Dore, O., 2007, PhRvD, 76, 043510\\
Spergel, D.~N. et al., 2007, ApJS, 170, 377\\
Staniszewski, Z., et al., 2009, \apj, 701, 32S\\
Sunyaev, R. A., \&  Zel'dovich, Ya. B., 1972, Comments Astrophys. Space Phys., 4, 173\\
Tegmark, M., \& de Oliveira-Costa, A., 1998, ApJ, 500, L83\\
Voit, G. M., \& Donahue, M., 2005, \apj, 634, 955\\
Weinmann, S. M., vandenBosch, F. C., Yang, X., Mo, H. J., Croton, D. J., \& Moore, B., 2006, \mnras, 372, 1161\\
White, M., Hernquist, L., Springel, V., 2002, \apj, 579, 16\\
Wright, E., et al.\, 2009, ApJS, 180, 283\\
Yamada, M., Sugiyama, N., \& Silk, J. 1999, \apj, 522, 66\\

\end{document}